\documentclass[a4paper,11pt]{article}
\pdfoutput=1

\usepackage{jheppub}
\usepackage[T1]{fontenc}

\usepackage{physics}
\usepackage{amssymb}

\usepackage{subcaption}
\usepackage{enumitem}

\allowdisplaybreaks

\renewcommand{\d}[2][]{\mathrm{d}^{#1}{#2}}
\newcommand{\de}[3][]{\frac{\mathrm{d}^{#1}{#2}}{\mathrm{d}{#3}^{#1}}}
\newcommand{\vol}{\mathrm{vol}}

\usepackage{letltxmacro}
\makeatletter
\let\oldr@@t\r@@t
\def\r@@t#1#2{%
\setbox0=\hbox{$\oldr@@t#1{#2\,}$}\dimen0=\ht0
\advance\dimen0-0.2\ht0
\setbox2=\hbox{\vrule height\ht0 depth -\dimen0}%
{\box0\lower0.4pt\box2}}
\LetLtxMacro{\oldsqrt}{\sqrt}
\renewcommand*{\sqrt}[2][\ ]{\oldsqrt[#1]{#2}}
\makeatother

\title{Complex Saddles and Euclidean Wormholes in the Lorentzian Path Integral
}
\author{Gregory J.\ Loges,}
\author{Gary Shiu}
\author{and Nidhi Sudhir}
\affiliation{Department of Physics, University of Wisconsin-Madison,\\1150 University Ave, Madison, WI 53706, USA}

\emailAdd{gloges@wisc.edu}
\emailAdd{shiu@physics.wisc.edu}
\emailAdd{kandathpatin@wisc.edu}

\abstract{
We study complex saddles of the Lorentzian path integral for 4D axion gravity and its dual description in terms of a 3-form flux, which include the Giddings-Strominger Euclidean wormhole. Transition amplitudes are computed using the Lorentzian path integral and with the help of Picard-Lefschetz theory. The number and nature of saddles is shown to qualitatively change in the presence of a bilocal operator that could arise, for example, as a result of considering higher-topology transitions. We also analyze the stability of the Giddings-Strominger wormhole in the 3-form picture, where we find that it represents a perturbatively stable Euclidean saddle of the gravitational path integral.
This calls into question the ultimate fate of such solutions in an ultraviolet-complete theory of quantum gravity.
}

\begin{document}

\maketitle

\section{Introduction}
\label{sec:intro}

Euclidean quantum gravity has proven quite powerful in understanding non-perturbative and thermodynamic aspects of gravitational systems (e.g.\ see \cite{Gibbons:1994cg}). For non-gravitational theories, Wick rotation transforms an oscillatory integral in Lorentzian signature into an exponentially suppressed integral in Euclidean signature. However, the same effect on convergence of path integrals is not guaranteed for gravitational theories as the Euclidean action for gravity is not positive definite. Despite various potential difficulties, including the conformal factor problem~\cite{Gibbons:1978ac}, it remains a quite useful tool in understanding black hole thermodynamics, quantum cosmology and AdS/CFT.

The question of which (if any) complex metrics should contribute to the gravitational path integral remains a question of great importance. The recent ``allowability'' criterion of~\cite{Kontsevich2021} (see also~\cite{Witten2021,Lehners:2021mah}) provides a simple diagnostic identifying to which complex metrics all possible $p$-form gauge theories may be consistently coupled. It remains to be seen if this is a necessary or sufficient criterion for consistently coupling any matter theory to gravity. There is also evidence that complex saddle points are \emph{required} to make sense of pure 3D gravity~\cite{Maloney:2007ud}. One approach is to avoid performing the Wick rotation to Euclidean signature at all and aim to understand the gravitational path integral from a Lorentzian perspective~\cite{Marolf:1996gb, Dasgupta:2001ue, Feldbrugge:2017fcc, DiazDorronsoro:2017hti,Brown:2017wpl, Marolf:2020rpm, Marolf:2021ghr}. We will adopt this approach as well and consider Lorentzian path integrals, using Picard-Lefschetz theory in the particularly visualizable case of a single complex variable to define oscillatory integrals. This provides a consistent framework in which Lorentzian, complex and Euclidean saddle points may be treated democratically, and has previously been fruitfully applied in the study of quantum cosmology and beyond~\cite{Cherman:2014sba,Tanizaki:2014xba, Behtash:2015loa,Feldbrugge:2017kzv,Feldbrugge:2017mbc,Feldbrugge:2017fcc,DiazDorronsoro:2017hti,Brown:2017wpl,Feldbrugge:2019fjs,Rajeev:2021yyl}.

In axion gravity the Giddings-Strominger (GS) wormhole~\cite{Giddings:1989bq} appears as a highly symmetric saddle point. Determining whether the GS wormhole represents a true minimum of the Euclidean action is critical in understanding its role and interpretation in the gravitational path integral. The nature of the GS critical point can be analyzed by considering fluctuations around the background geometry and looking for tachyonic directions in which the Euclidean action is decreased. There has been previous work in analyzing this saddle point in the axion picture~\cite{Hertog:2018kbz} (see also~\cite{Rubakov:1996cn, Kim:2003js, Alonso:2017avz, Hertog:22toappear}); here we approach the question of perturbative stability in the 3-form picture, which we argue provides a more transparent application of the correct boundary conditions. We find that no negative modes are present and the GS wormhole represents a true minimum of the Euclidean path integral.

Our findings have some bearing on the existence of Euclidean wormholes in string theory. Previous analysis of perturbative stability \cite{Hertog:2018kbz,Rubakov:1996cn,Kim:2003js} cast doubts on them being  genuine saddle points of the path integral. While such analyses have so far only been carried out for the GS wormhole, it was argued in \cite{Hertog:2018kbz} that any perturbative instabilty may explain away the puzzle with embedding Euclidean wormholes into AdS/CFT \cite{Maldacena:2004rf, Arkani-Hamed:2007cpn}. The correlation functions across the two boundary CFTs should factorize while the Euclidean wormhole saddle contributions (if they exist) would seem to indicate that they do not. Extending our work to include one or more dilaton would be necessary to establish the existence of Euclidean wormhole saddles in string theory (where the axions are accompanied by their dilaton partners). Our results, if they can be successfully extrapolated, would seem to suggest that the difficulty may not be in finding Euclidean wormhole saddles but in embedding them into AdS/CFT~\cite{Arkani-Hamed:2007cpn, Hertog:2017owm, Astesiano:2022qba, Marolf:2021kjc}. We note also in passing that establishing the existence of axionic Euclidean wormhole saddles is of importance to the Weak Gravity conjecture for axions~\cite{Arkani-Hamed:2006emk, Rudelius:2015xta, Brown:2015iha, Brown:2015lia, Montero:2015ofa, Heidenreich:2015nta, Hebecker:2016dsw, Hebecker:2018ofv, Andriolo:2020lul}.

The remainder of this paper is organized as follows. In Sec.~\ref{sec:duality} we review the duality between a 2-form gauge field/3-form flux and axion in 4D, emphasizing how the duality prescribes corresponding boundary conditions for the fields. In Sec.~\ref{sec:pathintegrals} we compute Lorentzian transition amplitudes on either side of the duality in saddle-point approximation and demonstrate how the analysis can be qualitatively different in the presence of a bilocal operator. Picard-Lefschetz theory is used to make sense of any oscillatory integrals encountered. In Sec.~\ref{sec:stability} we address the perturbative stability of the Giddings-Strominger wormhole in the 3-form picture. Finally, we conclude in Sec.~\ref{sec:disc}.


\section{Review of the duality}
\label{sec:duality}

The two systems we consider are axion coupled to Einstein gravity and a 2-form gauge field coupled to Einstein gravity, which are dual to one another in 4D. In this section we review this duality; the familiar reader can jump straight to Sec.~\ref{sec:pathintegrals}. See~\cite{Collinucci:2005opa} for a nice discussion of these ideas in Euclidean signature.

The duality is most clearly seen by going to a first-order formalism where the 3-form flux is the dynamical field. Write
\begin{equation}\label{eq:gHth_action}
\begin{aligned}
    S[g,H,\theta] &= S_\text{grav}[g] + S_\text{m}[g,H,\theta] \,,\\
    S_\text{grav}[g] &= \frac{1}{2}\int_\mathcal{M}\d[4]{x}\,\sqrt{-g}\,R + \epsilon\int_{\partial\mathcal{M}}\d[3]{x}\,\sqrt{|h|}\,K \,,\\
    S_\text{m}[g,H,\theta] &= \int_\mathcal{M}\left(-\frac{1}{2}H\wedge{\star H} + \theta\,\d{H}\right) \,,
\end{aligned}
\end{equation}
where $\epsilon=\pm1$ for time-like/space-like boundaries. The variational problem for the above action is well-posed if one chooses Dirichlet boundary conditions for both the metric (due to the Gibbons-Hawking-York term) and 3-form (restricting to a particular choice of flux). There are no boundary conditions on the Lagrange multiplier $\theta$ which has been introduced to impose $\d{H}=0$ as a constraint. Path integrals on a fixed manifold $\mathcal{M}$ take the form
\begin{equation}
    Z_\mathcal{M}[h,J] = \int_{g|_\partial=h,\;H|_\partial = J}\hspace{-50pt}\mathcal{D}g\mathcal{D}H\mathcal{D}\theta\;e^{iS[g,H,\theta]} \,,
\end{equation}
where $h$, $J$ indicate the chosen boundary conditions for $g$, $H$ respectively. Of course, with dynamical gravity one can imagine specifying only the boundary manifold and performing a sum over bulk manifolds with the specified boundary:
\begin{equation}
    Z_\mathcal{B}[h,J] = \sum_{\mathcal{M}\;:\;\partial\mathcal{M}=\mathcal{B}}Z_\mathcal{M}[h,J] \,.
\end{equation}
Such a topological expansion has been well-explored in the context of 2D dilaton gravity, e.g.\ in Jackiw-Teitelboim gravity~\cite{Saad:2019lba}.

On one hand, the Lagrange multiplier may be integrated out exactly, resulting in the constraint that $H$ is closed (i.e.\ $H=\d{B}$ locally) and the action for a 2-form gauge field. On the other, the 3-form flux may be integrated out by completing the square in $H$:
\begin{equation}
    S_\text{m}[g,H,\theta] = \int_\mathcal{M}\left(-\frac{1}{2}(H-{\star\d{\theta}})\wedge{\star(H-{\star\d{\theta}})} - \frac{1}{2}\d{\theta}\wedge{\star\d{\theta}}\right) + \int_{\partial\mathcal{M}}\theta H \,.
\end{equation}
We have used that $\star^2=(-1)^{p+1}$ for 4D Lorentzian manifolds when acting on $p$-forms. The integral over the bulk is Gaussian and $H|_\partial = J$ is fixed as a boundary condition. This allows us to write
\begin{equation}
\begin{aligned}
    Z_\mathcal{M}[h,J] &= \int_{g|_\partial = h}\hspace{-20pt}\mathcal{D}g\mathcal{D}\theta\;e^{iS[g,\theta]} \,,\\
    S[g,\theta] &= S_\text{grav}[g] + \int_\mathcal{M}\left(-\frac{1}{2}\d{\theta}\wedge{\star\d{\theta}}\right) + \int_{\partial\mathcal{M}}\theta J \,.
\end{aligned}
\end{equation}
If we split the axion into bulk and boundary degrees of freedom then the above has the structure of a Fourier transform:
\begin{equation}
\begin{aligned}
    Z_\mathcal{M}[h,J] &= \int\mathcal{D}\theta_\text{bdy}\;e^{i\int_{\partial\mathcal{M}}\theta_\text{bdy} J}\int_{g|_\partial = h,\;\theta|_\partial = \theta_\text{bdy}}\hspace{-50pt}\mathcal{D}g\mathcal{D}\theta \;e^{iS_\text{grav}[g] + i\int_\mathcal{M}\big(-\frac{1}{2}\d{\theta}\wedge{\star\d{\theta}}\big)} \,,\\
    &\equiv \int\mathcal{D}\theta_\text{bdy}\;e^{i\int_{\partial\mathcal{M}}\theta_\text{bdy} J} Z_\mathcal{M}[h,\theta_\text{bdy}] \,.
\end{aligned}
\end{equation}
Alternatively, one can view this as imposing Neumann boundary conditions on the axion.

To summarize, the path integral $Z_\mathcal{M}[h,J]$ may be computed in one of two ways. In the 3-form picture one simply computes
\begin{equation}
    Z_\mathcal{M}[h,J] = \int_{g|_\partial=h,\;H|_\partial = J}\hspace{-50pt}\mathcal{D}g\mathcal{D}H\;\delta[\d{H}]\;e^{iS_\text{grav}[g] + i\int_\mathcal{M}\big(-\frac{1}{2}H\wedge{\star H}\big)} \,,
\end{equation}
while in the axion picture one can structure the computation as
\begin{equation}
\begin{aligned}
    Z_\mathcal{M}[h,J] &= \int\mathcal{D}\theta_\text{bdy}\;e^{i\int_{\partial\mathcal{M}}\theta_\text{bdy} J} Z_\mathcal{M}[h,\theta_\text{bdy}] \,,\\
    Z_\mathcal{M}[h,\theta_\text{bdy}] &= \int_{g|_\partial = h,\;\theta|_\partial = \theta_\text{bdy}}\hspace{-50pt}\mathcal{D}g\mathcal{D}\theta \;e^{iS_\text{grav}[g] + i\int_\mathcal{M}\big(-\frac{1}{2}\d{\theta}\wedge{\star\d{\theta}}\big)} \,.
\end{aligned}
\end{equation}
In the next section we will compute one such $Z_\mathcal{M}[h,J]$ in some detail on either side of this duality.


\section{Lorentzian path integrals}
\label{sec:pathintegrals}


\subsection{Transition amplitudes}
\label{sec:amplitudes}

In this section we will examine transition amplitudes between two boundaries of $S^3$ topology, restricting attention to $\mathcal{M}=\mathbb{R}\times S^3$ as the leading contribution in a topological expansion. We can take the coordinate time to be $t\in[0,1]$ so that the boundaries lie at $t=0$ and $t=1$. For boundary conditions we take
\begin{equation}
\begin{aligned}
    \d{s^2}\big|_{t=0} &= q_0\,\d{\Omega_3^2} \,, & \qquad J\big|_{t=0} &= \frac{n_0\,\vol_3}{2\pi^2} \,,\\
    \d{s^2}\big|_{t=1} &= q_1\,\d{\Omega_3^2} \,, & J\big|_{t=1} &= \frac{n_1\,\vol_3}{2\pi^2} \,,
\end{aligned}
\end{equation}
where $\d{\Omega_3^2}$ is the round metric on $S^3$ and $\vol_3$ is the corresponding volume form normalized to $\int_{S^3}\vol_3 = 2\pi^2$. Let's begin with the 3-form computation; without loss of generality, we will consider only $q_1\geq q_0\geq 0$ and work with the following homogeneous, isotropic ansatz
\begin{equation}\label{eq:metricHansatz}
\begin{aligned}
    \d{s^2} &= -\frac{N^2}{q(t)}\,\d{t^2} + q(t)\,\d{\Omega_3^2} \,,\\
    H &= \mathfrak{h}(t)\,\vol_3 \,,
\end{aligned}
\end{equation}
with $N$ constant. The transition amplitude of interest is then
\begin{equation}
\begin{aligned}
    Z[q_0,q_1;n_0,n_1] &\simeq \int_0^\infty\d{N}\int_{q(0)=q_0}^{q(1)=q_1}\mathcal{D}q\int_{\mathfrak{h}(0)=n_0}^{\mathfrak{h}(1)=n_1}\mathcal{D}\mathfrak{h}\;\delta[\dot{\mathfrak{h}}(t)]\;e^{iS[N,q,\mathfrak{h}]} \,,\\
    S[N,q,\mathfrak{h}] &= 2\pi^2\int_0^1\d{t}\,\left( 3N - \frac{3\dot{q}^2}{4N} - \frac{\mathfrak{h}^2N}{2q^2} \right) \,,
\end{aligned}
\end{equation}
where $\dot{A}\equiv\de{A}{t}$. The path integral over $\mathfrak{h}$ may immediately be done since the constraint $\d{H}=0$ fixes $\dot{\mathfrak{h}}=0$. The transition amplitude is nonzero only if $n_0=n_1\equiv n$, in which case one simply has $\mathfrak{h}(t)=\frac{n}{2\pi^2}$ with $n$ integer-quantized:
\begin{equation}
    n = \int_{S^3}H \in\mathbb{Z} \,.
\end{equation}
To reduce notational clutter it is convenient to introduce the rescaled flux $\tilde{n}=\frac{n}{2\pi^2\sqrt{6}}$ and restrict attention to $\tilde{n}>0$. Stripping off the flux-conserving $\delta$-function, we have
\begin{equation}\label{eq:Hpath}
    Z[q_0,q_1;n] \simeq \int_0^\infty\d{N}\int_{q(0)=q_0}^{q(1)=q_1}\mathcal{D}q\;e^{iS[N,q,n]} \,.
\end{equation}
Our approach will be to evaluate the $q$ path integral for any boundary conditions $q_0,q_1$ and then use Picard-Lefschetz theory to evaluate the remaining one-dimensional integral over $N$. The equation of motion for $q$ is
\begin{equation}
    \ddot{q} + \frac{4\tilde{n}^2N^2}{q^3} = 0 \,,
\end{equation}
for which there are two solutions satisfying the boundary conditions,
\begin{equation}
    q_\pm(t) = \sqrt{q_0^2(1-t)^2 + q_1^2t^2 + 2q_0q_1\Gamma_\pm\,t(1-t)} \,, \qquad \Gamma_\pm = \pm\sqrt{1 + \frac{4\tilde{n}^2N^2}{q_0^2q_1^2}} \,,
\end{equation}
where we take the principle branch of the square-root so that $q_\pm(0)=\sqrt{q_0^2}=q_0$ and $q_\pm(1)=\sqrt{q_1^2}=q_1$. It is straightforward to show that $q_+^2$ is nowhere zero on $t\in(0,1)$ whereas $q_-^2$ always passes through zero. Consequently, we take only $q(t)=q_+(t)$ as a viable classical saddle point. One finds
\begin{equation}
    S[N,q_+,n] = 2\pi^2\left[3N - \frac{3(q_0^2+q_1^2)}{4N}+\frac{3q_0q_1}{2N}\,f\Big(\frac{2\tilde{n}N}{q_0q_1}\Big)\right] \,,
\end{equation}
where
\begin{equation}
    f(z) = \sqrt{z-i}\sqrt{z+i} - z\log{\big(z+\sqrt{z-i}\sqrt{z+i}\big)} \,.
\end{equation}
There are branch points at $z=\pm i$ and we have chosen the branch cuts to extend parallel to the negative real axis for sake of presentation (see Fig.~\ref{fig:PL_H}). We arrive at
\begin{equation}
    Z[q_0,q_1;n] \simeq \int_0^\infty\d{N}\,e^{2\pi^2i\left[3N - \frac{3(q_0^2+q_1^2)}{4N}+\frac{3q_0q_1}{2N}f\big(\frac{2\tilde{n}N}{q_0q_1}\big) \right]} \,,
\end{equation}
and we evaluate this oscillatory function using Picard-Lefschetz theory. There are four saddle points of $S[N,q_+,n]$, which occur where
\begin{equation}
    16N^4 - 8N^2(2\tilde{n}^2-q_0^2-q_1^2) + (q_0^2-q_1^2)^2 = 0 \,,
\end{equation}
namely at
\begin{equation}\label{eq:saddles}
    N_{\pm\pm} = \frac{1}{2}\left(\pm\sqrt{\tilde{n}^2 - q_0^2} \pm \sqrt{\tilde{n}^2 - q_1^2}\right)
\end{equation}
with the two signs being chosen independently. The initial contour $N\in(0,\infty)$ is deformed to intersect one or more of these saddle points in such a way that $\Im(iS)$ is (piecewise) constant and $\Re(iS)$ decreases as one moves on the contour away from a saddle -- this ensures that the resulting integrals are absolutely convergent. The required contour deformation depends on the relative values of $q_0,q_1,\tilde{n}$ and falls into the following three cases:

\begin{figure}[t]
    \centering
    \vspace{35pt}
    \begin{subfigure}{0.49\textwidth}
        \centering
        \includegraphics[width=\linewidth]{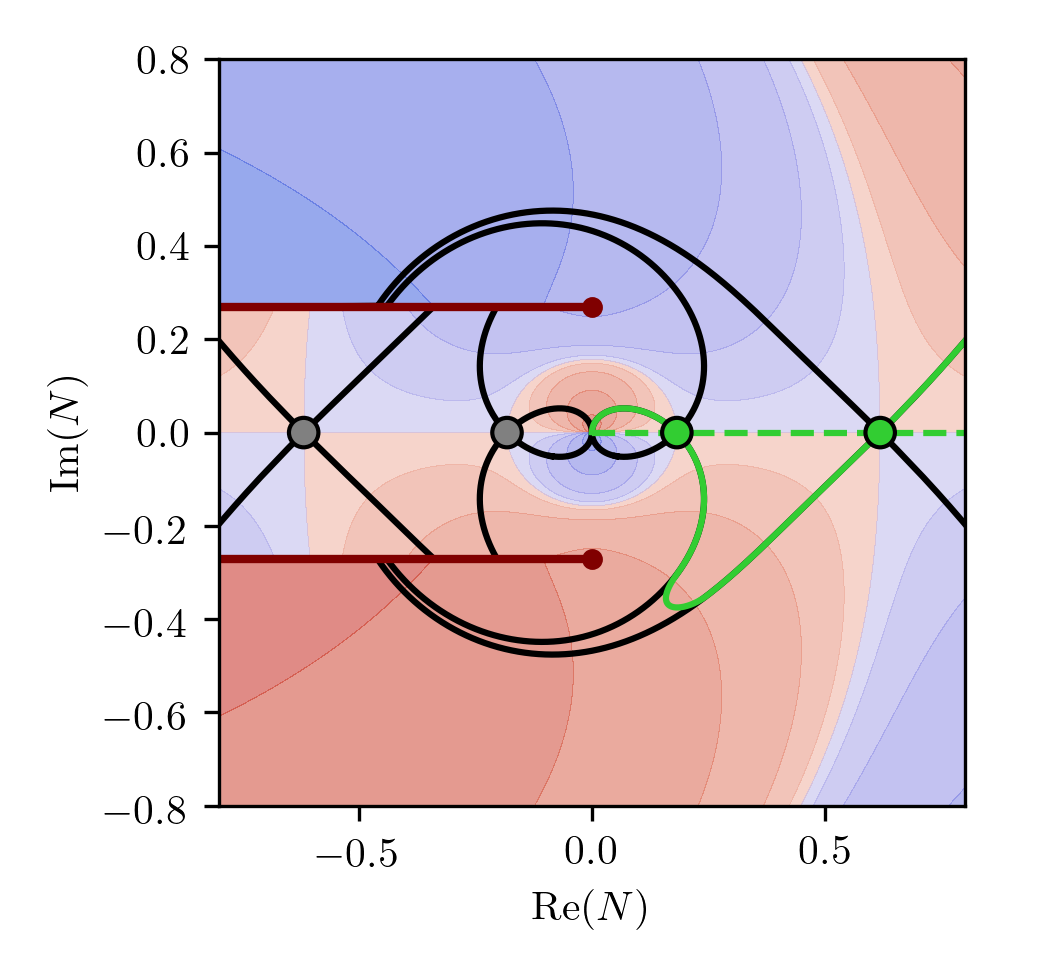}
        \subcaption{$(q_0,q_1,\tilde{n})=(0.6,0.9,1)$}
        \label{fig:PL_H_1}
    \end{subfigure}\hfill
    \begin{subfigure}{0.49\textwidth}
        \centering
        \includegraphics[width=\linewidth]{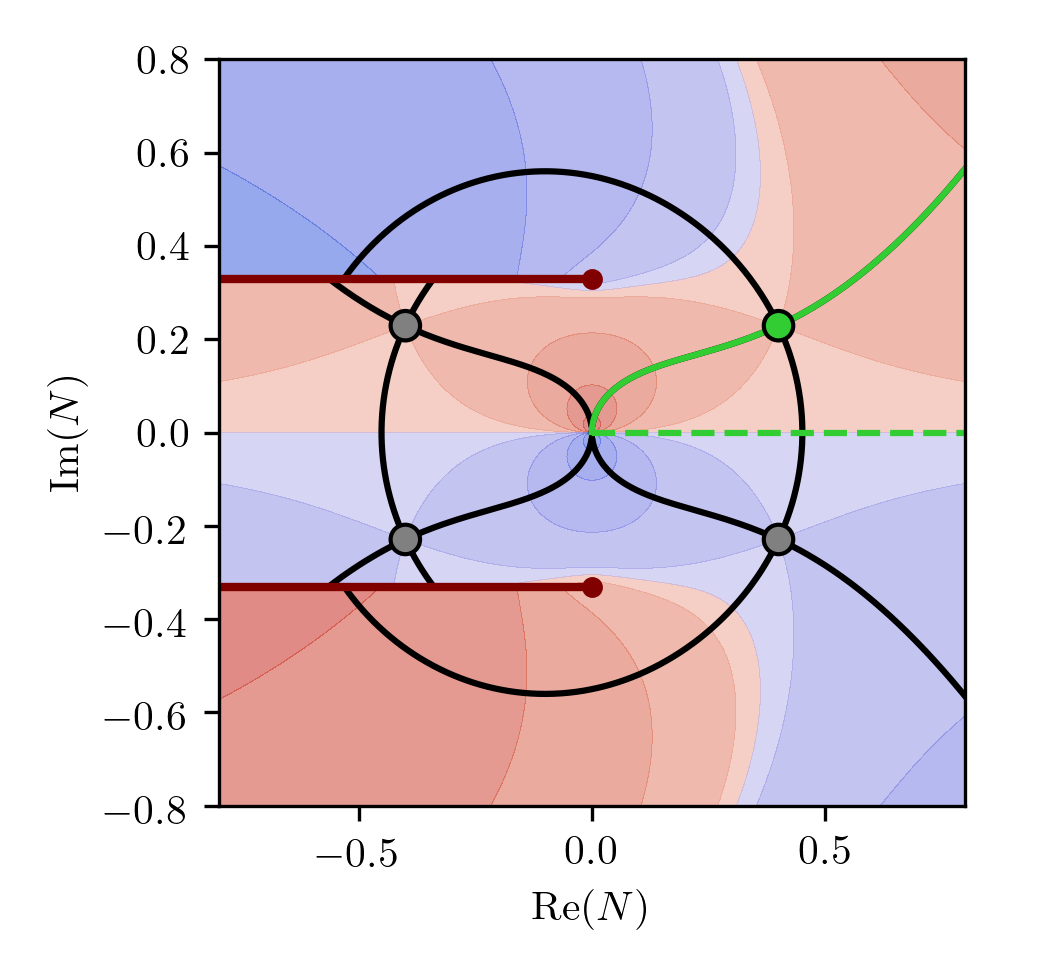}
        \subcaption{$(q_0,q_1,\tilde{n})=(0.6, 1.1, 1)$}
        \label{fig:PL_H_2}
    \end{subfigure}\\
    \begin{subfigure}{0.49\textwidth}
        \centering
        \includegraphics[width=\linewidth]{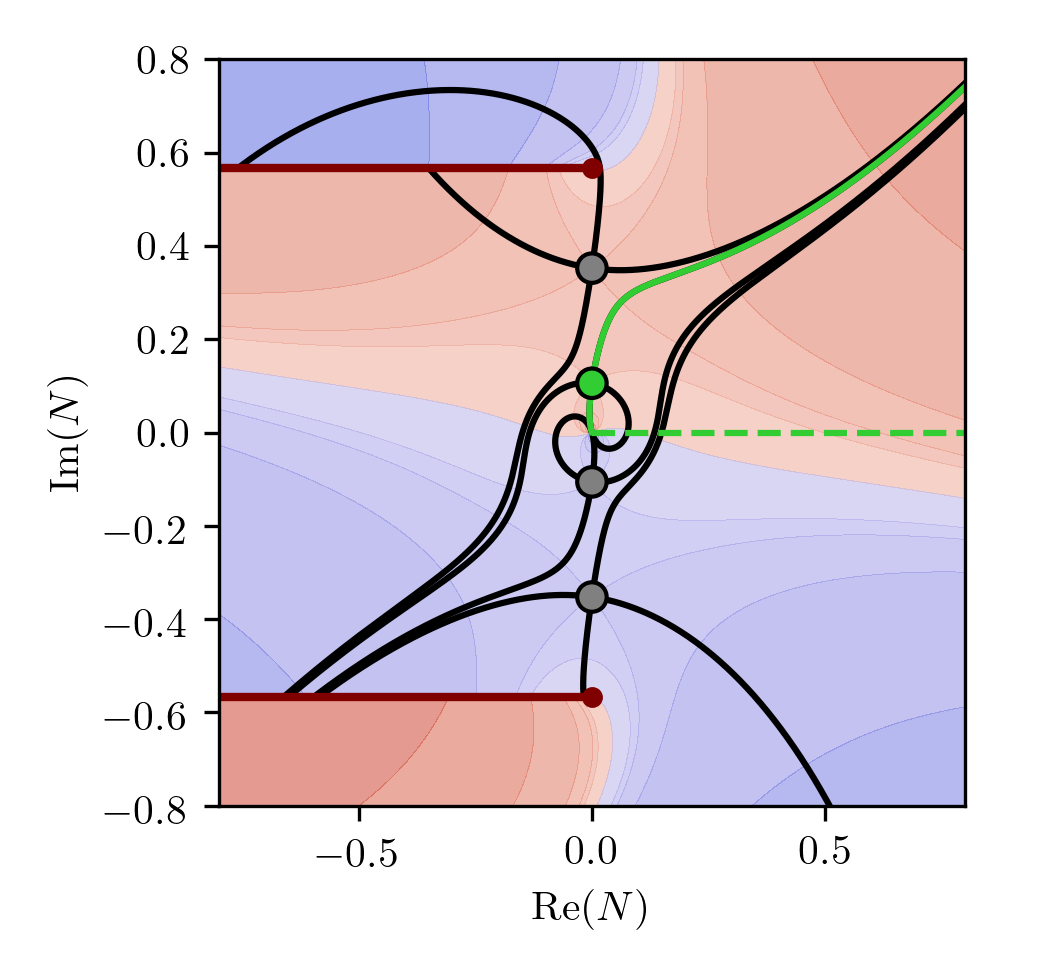}
        \subcaption{$(q_0,q_1,\tilde{n})=(1.03, 1.1, 1)$ with $\hbar=1-0.3i$}
        \label{fig:PL_H_3}
    \end{subfigure}\hfill
    \begin{subfigure}{0.49\textwidth}
        \centering
        \includegraphics[width=\linewidth]{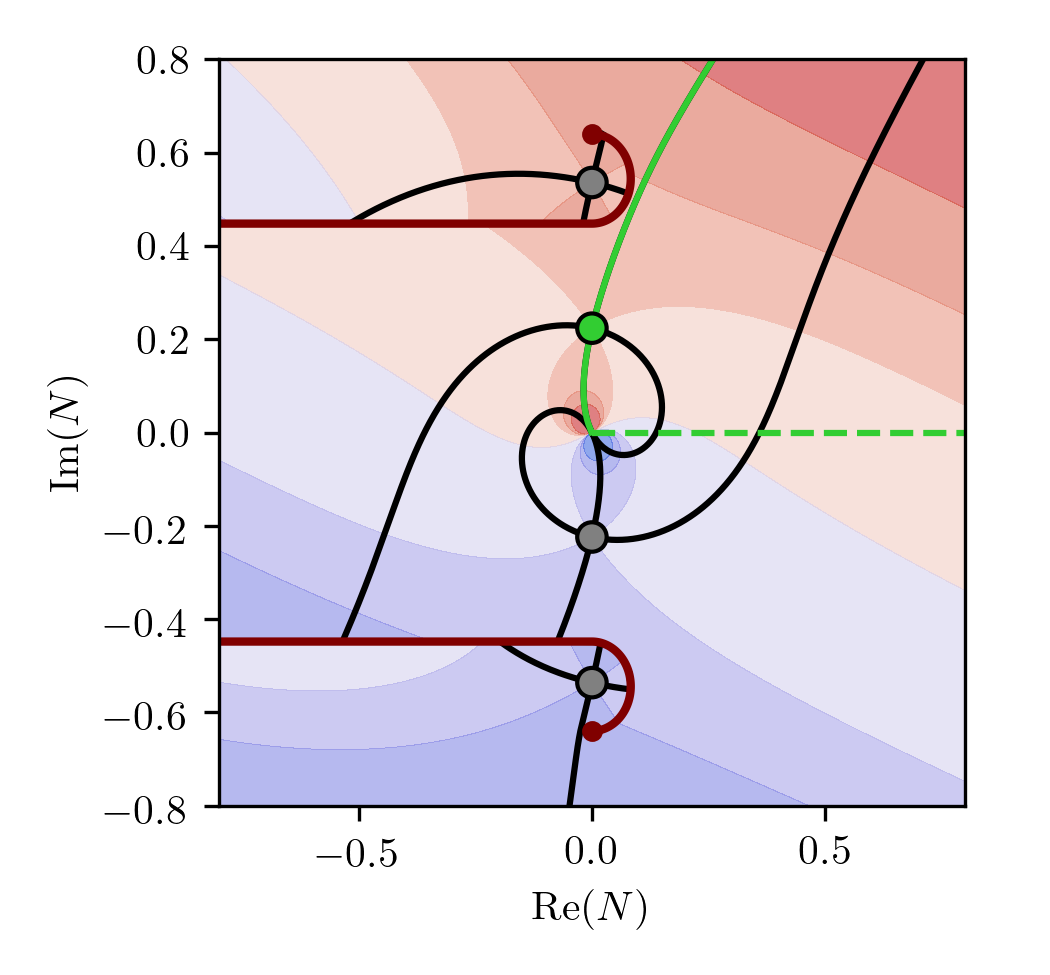}
        \subcaption{$(q_0,q_1,\tilde{n})=(0.4, 0.8, 0.25)$ with $\hbar=1-0.5i$}
        \label{fig:PL_H_4}
    \end{subfigure}
    \caption{Structure of $iS[N,q_+,n]$ in the complex-$N$ plane for (a) $q_0,q_1<\tilde{n}$, (b) $q_0<\tilde{n}<q_1$, (c) $q_0,q_1>\tilde{n}$ with $\frac{1}{q_0}+\frac{1}{q_1}>\frac{1}{\tilde{n}}$ and (d) $q_0,q_1>\tilde{n}$ with $\frac{1}{q_0}+\frac{1}{q_1}<\frac{1}{\tilde{n}}$. In (c) and (d) we have set $\Im(\hbar)<0$ to resolve ambiguities in the Picard-Lefschetz prescription as described in the main text. Regions of red have $\Re(iS)<0$ and regions of blue have $\Re(iS)>0$: several level-sets of $\Re(iS)$ are shown with dotted lines. Steepest ascent/descent contours (level-sets of $\Im(iS)$ emanating from the saddle points) are shown with solid black lines. The initial and deformed $N$-contours are shown with dashed and solid green lines, respectively.\vspace{35pt}}
    \label{fig:PL_H}
\end{figure}

\begin{enumerate}[label=\roman*)]
    \item $q_0< q_1<\tilde{n}$: all saddles are on the real axis and the deformed contour intersects the two positive saddle points with angles $\pm45^\circ$ (see Fig.~\ref{fig:PL_H_1}).
    \item $q_0<\tilde{n}<q_1$: all saddles are complex and the deformed contour intersects the one saddle point in the positive quadrant (see Fig.~\ref{fig:PL_H_2}).
    \item $\tilde{n}<q_0<q_1$: all saddles are on the imaginary axis and the steepest ascent/descent contours intersect the branch points of $S[N,q_+,n]$. To resolve this ambiguity we shift $S\to \frac{S}{\hbar}=\frac{S}{1-i\epsilon}$ and then take $\epsilon\to0^+$. The deformed contour intersects one saddle point. For $\frac{1}{q_0}+\frac{1}{q_1}= \frac{1}{\tilde{n}}$ two of the four saddles collide with the branch points and for $\frac{1}{q_0}+\frac{1}{q_1} < \frac{1}{\tilde{n}}$ they move to another sheet; however, these saddle points are never included under the $\epsilon$-deformation and so this curiosity will not concern us (see Fig.~\ref{fig:PL_H_3} and Fig.~\ref{fig:PL_H_4}).
\end{enumerate}
Having identified which saddles contribute to the transition amplitude, we may then evaluate the transition amplitude in each of the three cases above. These are compactly written in terms of
\begin{equation}
\begin{aligned}
    g_1(z) &= \sqrt{1-z^2} - \operatorname{arcsech}{z} & z\leq1\\
    g_2(z) &= ig_1(z+i\epsilon) = \sqrt{z^2-1} - \arcsec{z} & \quad z\geq1
\end{aligned}
\end{equation}
as
\begin{equation}\label{eq:Z3form}
    Z[q_0,q_1;n] \simeq \begin{cases}
        e^{-i\pi/4}e^{6\pi^2i\tilde{n}[g_1(\frac{q_0}{\tilde{n}}) - g_1(\frac{q_1}{\tilde{n}})]} + e^{i\pi/4}e^{6\pi^2i\tilde{n}[g_1(\frac{q_0}{\tilde{n}}) + g_1(\frac{q_1}{\tilde{n}})]} & q_0 < q_1<\tilde{n}\\
        e^{i\alpha}e^{6\pi^2i\tilde{n}g_1(\frac{q_0}{\tilde{n}})}e^{-6\pi^2\tilde{n}g_2(\frac{q_1}{\tilde{n}})} & q_0<\tilde{n}<q_1\\
        e^{i\pi/2}e^{-6\pi^2\tilde{n}[g_2(\frac{q_1}{\tilde{n}})-g_2(\frac{q_0}{\tilde{n}})]} & \tilde{n}<q_0<q_1
    \end{cases}
\end{equation}
where $\alpha$ gives the angle of the steepest-descent contours through the sole contributing saddle point -- an unilluminating function of $\frac{q_0}{\tilde{n}}$ and $\frac{q_1}{\tilde{n}}$. For $q_0,q_1<\tilde{n}$ there are two Lorentzian saddles which contribute as phases and lead to interference effects. For $q_0<\tilde{n}<q_1$ we can interpret the $q_0$-dependent phase as describing Lorentzian evolution from $q=q_0$ to $q=\tilde{n}$ and then the $q_1$-dependent exponential suppression as due to the tunnelling into the classically forbidden region $q>\tilde{n}$. For $\tilde{n}<q_0<q_1$ the saddle point is Euclidean and the bulk geometry is a segment of the full GS wormhole: the exponential suppression can be identified with its action.

\bigskip

\begin{figure}[t]
    \centering
    \begin{subfigure}{\textwidth}
        \centering
        \includegraphics[width=0.95\linewidth]{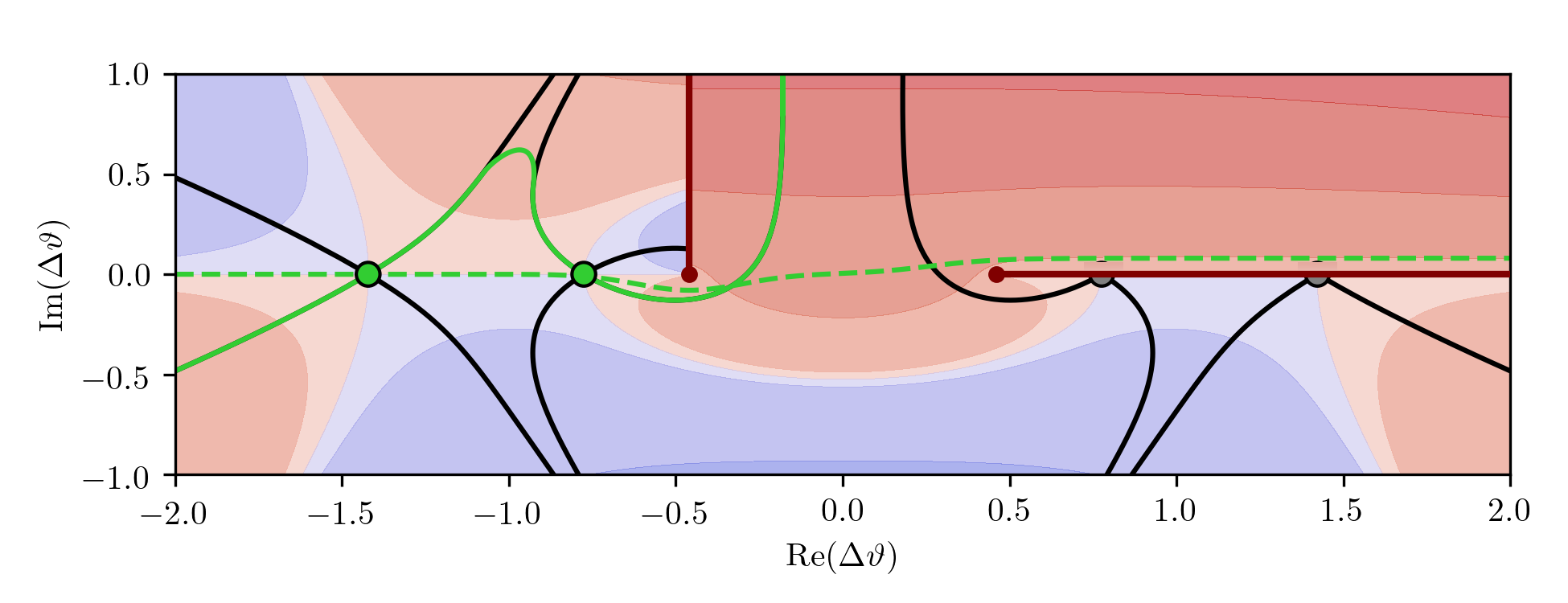}
        \subcaption{$(q_0,q_1,\tilde{n})=(0.6,0.95,1)$}
        \label{fig:PL_axion_1}
    \end{subfigure}
    \begin{subfigure}{0.49\textwidth}
        \centering
        \includegraphics[width=\linewidth]{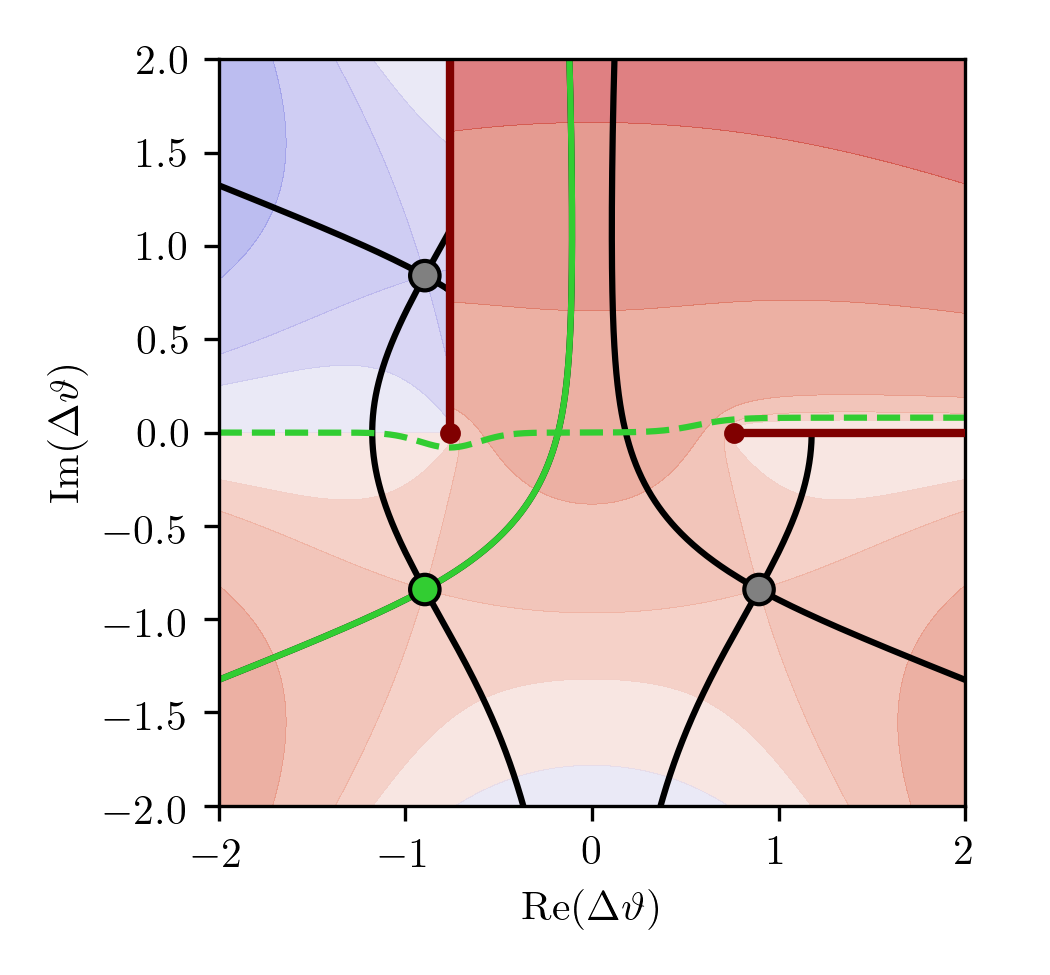}
        \subcaption{$(q_0,q_1,\tilde{n})=(0.7,1.5,1)$}
        \label{fig:PL_axion_2}
    \end{subfigure}\hfill
    \begin{subfigure}{0.49\textwidth}
        \centering
        \includegraphics[width=\linewidth]{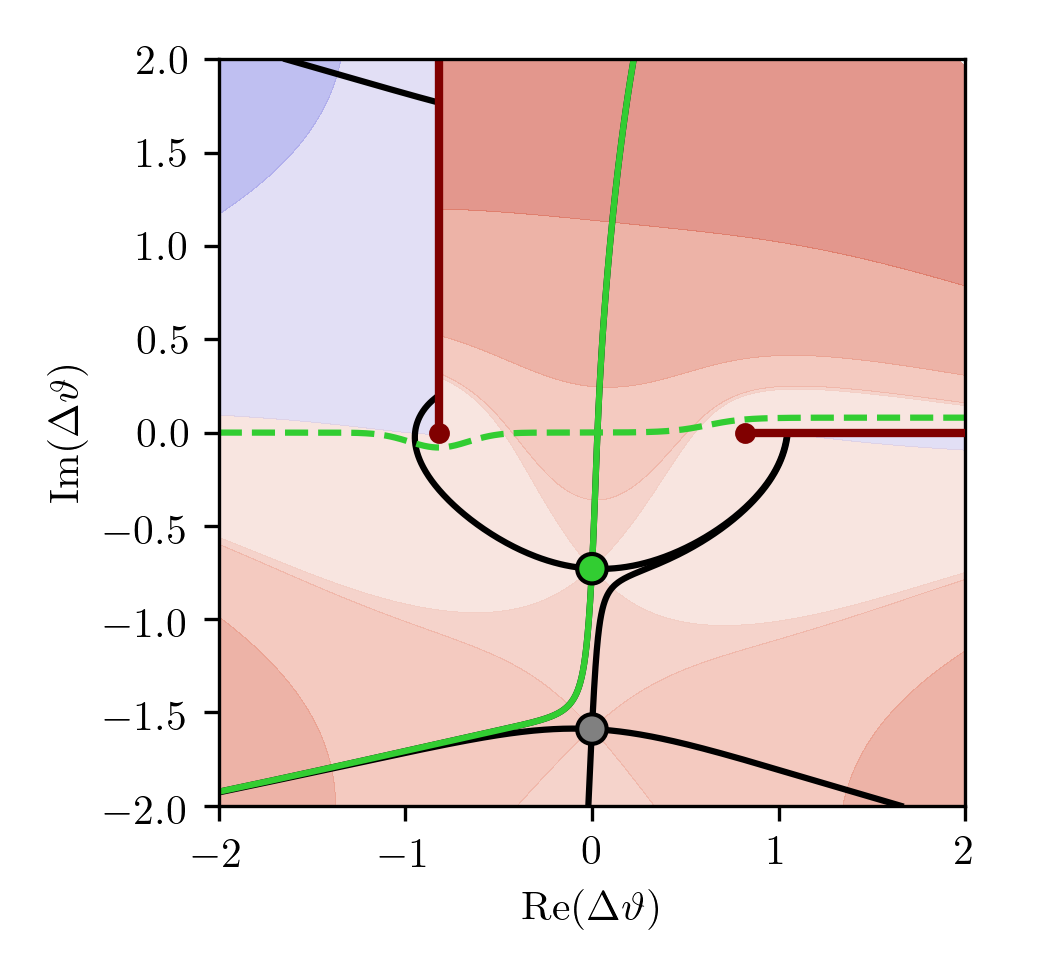}
        \subcaption{$(q_0,q_1,\tilde{n})=(1.1,2.5,1)$ with $\hbar=1-0.1i$}
        \label{fig:PL_axion_3}
    \end{subfigure}
    \caption{Structure of $\Phi$ of Eqn.~\eqref{eq:FT} in the complex-$\Delta\vartheta$ plane for (a) $q_0,q_1<\tilde{n}$, (b) $q_0<\tilde{n}<q_1$ and (c) $\tilde{n}<q_0<q_1$. In (c) we have set $\Im(\hbar)<0$ to resolve ambiguities in the Picard-Lefschetz prescription. Regions of red have $\Re(i\Phi)<0$ and regions of blue have $\Re(i\Phi)>0$: several level-sets of $\Re(i\Phi)$ are shown with dotted lines. Steepest ascent/descent contours are shown with solid black lines. The initial and deformed $\Delta\vartheta$-contours are shown with dashed and solid green lines, respectively.}
    \label{fig:PL_axion}
\end{figure}

Turning now to the axion side of the duality, we see that the restriction to $n_0=n_1$ arises due to the shift symmetry of the theory:
\begin{equation}
\begin{aligned}
    Z[q_0,q_1;n_0,n_1] &\simeq \int\d{\theta_0}\d{\theta_1} e^{i(\theta_1n_1-\theta_0n_0)}Z[q_0,q_1;\theta_0,\theta_1]\\
    &= \int\d{\overline{\theta}}\d{\Delta\theta}\,e^{i[\overline{\theta}(n_1-n_0) + \Delta\theta(\frac{n_0+n_1}{2})]}Z[q_0,q_1;0,\Delta\theta]\\
    &= 2\pi\,\delta(n_1-n_0)\int\d{\Delta\theta}\,e^{i\Delta\theta(n_0+n_1)/2}Z[q_0,q_1;0,\Delta\theta] \,.
\end{aligned}
\end{equation}
Calculating $Z[q_0,q_1;0,\Delta\theta]$ is relatively easy. We have
\begin{equation}
\begin{aligned}
    Z[q_0,q_1;0,\Delta\theta] &\simeq \int_0^\infty\d{N}\int_{q(0)=q_0}^{q(1)=q_1}\mathcal{D}q\int_{\theta(0)=0}^{\theta(1)=\Delta\theta}\mathcal{D}\theta\,e^{iS[N,q,\theta]} \,,\\
    S[N,q,\theta] &= 2\pi^2\int_0^1\d{t}\,\left(3N - \frac{3\dot{q}^2}{4N} + \frac{q^2\dot{\theta}^2}{2N}\right) \,,
\end{aligned}
\end{equation}
and the resulting equations of motion for $q$ and $\theta$ are (introducing $\vartheta=\sqrt{\frac{2}{3}}\theta$),
\begin{equation}
    \ddot{q} + q\dot{\vartheta}^2 = 0 \,, \qquad \de{}{t}\big(q^2\dot{\vartheta}\big) = 0 \,,
\end{equation}
which have solution
\begin{equation}
\begin{aligned}
    q(t) &= \sqrt{q_0^2(1-t)^2 + q_1^2t^2 + 2q_0q_1\cosh{(\Delta\vartheta)}\,t(1-t)} \,,\\
    \vartheta(t) &= \frac{1}{2}\log{\left( \frac{q_0(1-t)+q_1e^{\Delta\vartheta}t}{q_0(1-t)+q_1e^{-\Delta\vartheta}t} \right)} \,.
\end{aligned}
\end{equation}
Much like for the 3-form, there is another candidate solution satisfying the boundary conditions which is discarded because it results in $q^2<0$ in the interval $t\in(0,1)$. Because the above solution is independent of $N$, the remaining one-dimensional integral is particularly simple:
\begin{equation}
\begin{aligned}
    Z[q_0,q_1;0,\Delta\theta] &\simeq \int_0^\infty\d{N}\,e^{6\pi^2i\left(N-\frac{q_0q_1}{2N}\big[\cosh{\big(\log{\frac{q_1}{q_0}}\big)}-\cosh{(\Delta\vartheta)}\big]\right)}\\
    &= i\sqrt{b-i\epsilon}\,K_1\big(6\pi^2\sqrt{b-i\epsilon}\big) \,,
\end{aligned}
\end{equation}
where $b = 2q_0q_1\big[\cosh{\big(\log{\tfrac{q_1}{q_0}}\big)} - \cosh{(\Delta\vartheta)}\big]$ and $K_1(z)$ is a modified Bessel function of the second kind. For our purposes it will suffice to use the approximation
\begin{equation}
    K_1(z) \sim \sqrt{\frac{\pi}{2}}\,\frac{e^{-z}}{\sqrt{z}}
\end{equation}
to arrive at
\begin{equation}
    Z[q_0,q_1;0,\Delta\theta] \simeq \frac{i}{\sqrt{12\pi}}(b-i\epsilon)^{1/4}e^{-6\pi^2\sqrt{b-i\epsilon}} \simeq e^{-6\pi^2\sqrt{b-i\epsilon}} \,,
\end{equation}
where again we drop all prefactors. Finally, the remaining Fourier transform takes the form
\begin{equation}\label{eq:FT}
\begin{aligned}
    Z[q_0,q_1;n] &\simeq \int\d{\Delta\theta}\,e^{in\Delta\theta}e^{-6\pi^2\sqrt{b-i\epsilon}} \simeq \int\d{\Delta\vartheta}\,e^{i\Phi} \,,\\
    \Phi &= 6\pi^2\Big(\tilde{n}\Delta\vartheta + i\sqrt{2q_0q_1}\sqrt{\cosh{(\log{\tfrac{q_1}{q_0}})} - \cosh{(\Delta\vartheta)} - i\epsilon}\Big) \,.
\end{aligned}
\end{equation}
This oscillatory integral may be performed in saddle-point approximation using Picard-Lefschetz theory. One finds that the locations of the saddle-points in $\Delta\vartheta$ and thus the deformation of the initial $\Delta\vartheta$-contour falls into three cases exactly matching those found in the 3-form analysis: see Fig.~\ref{fig:PL_axion}. Evaluating Eqn.~\eqref{eq:FT} on the selected saddle(s) leads to perfect agreement with Eqn.~\eqref{eq:Z3form}, as expected.


\subsection{Higher topologies}
\label{sec:top}

\begin{figure}[h]
    \centering
    \begin{subfigure}{0.49\textwidth}
        \centering
        \includegraphics[width=\linewidth]{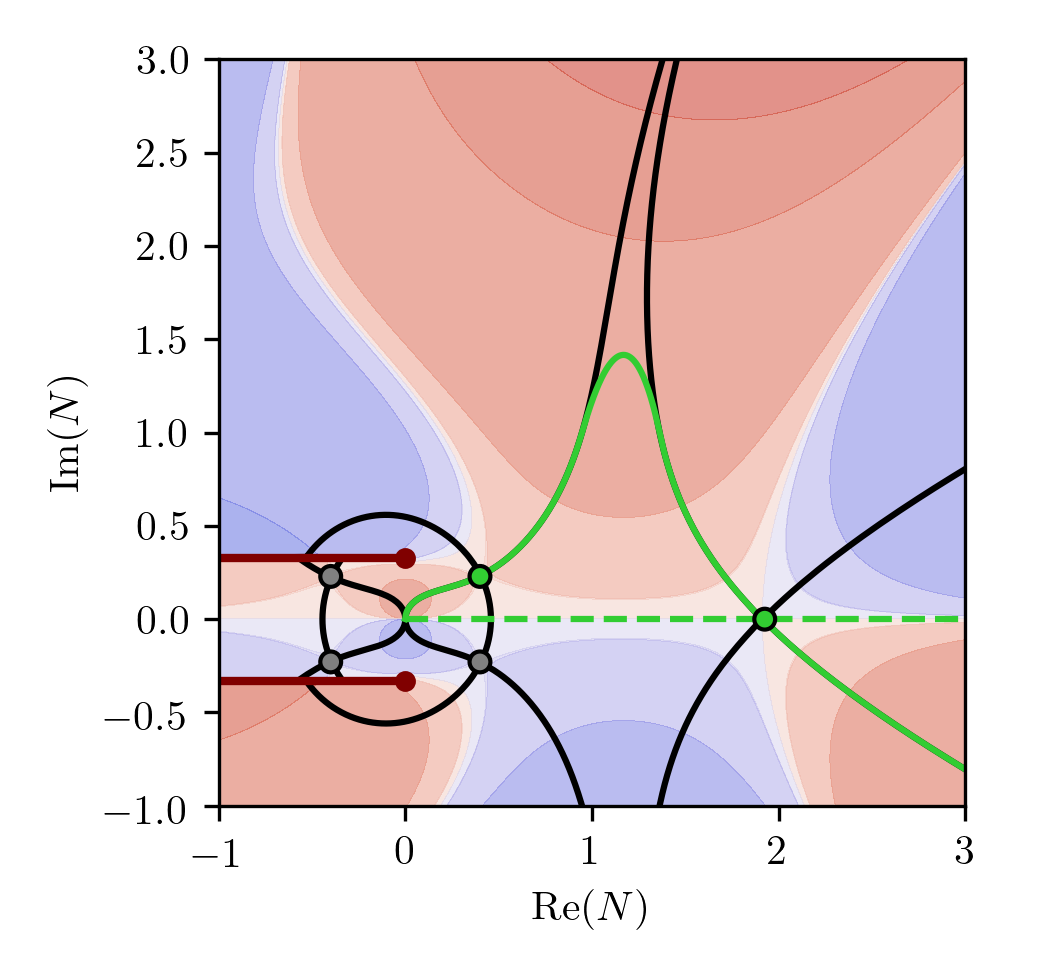}
        \subcaption{$(q_0,q_1,\tilde{n})=(0.6,1.1,1)$ and $c=0.1$}
    \end{subfigure}\hfill
    \begin{subfigure}{0.49\textwidth}
        \centering
        \includegraphics[width=\linewidth]{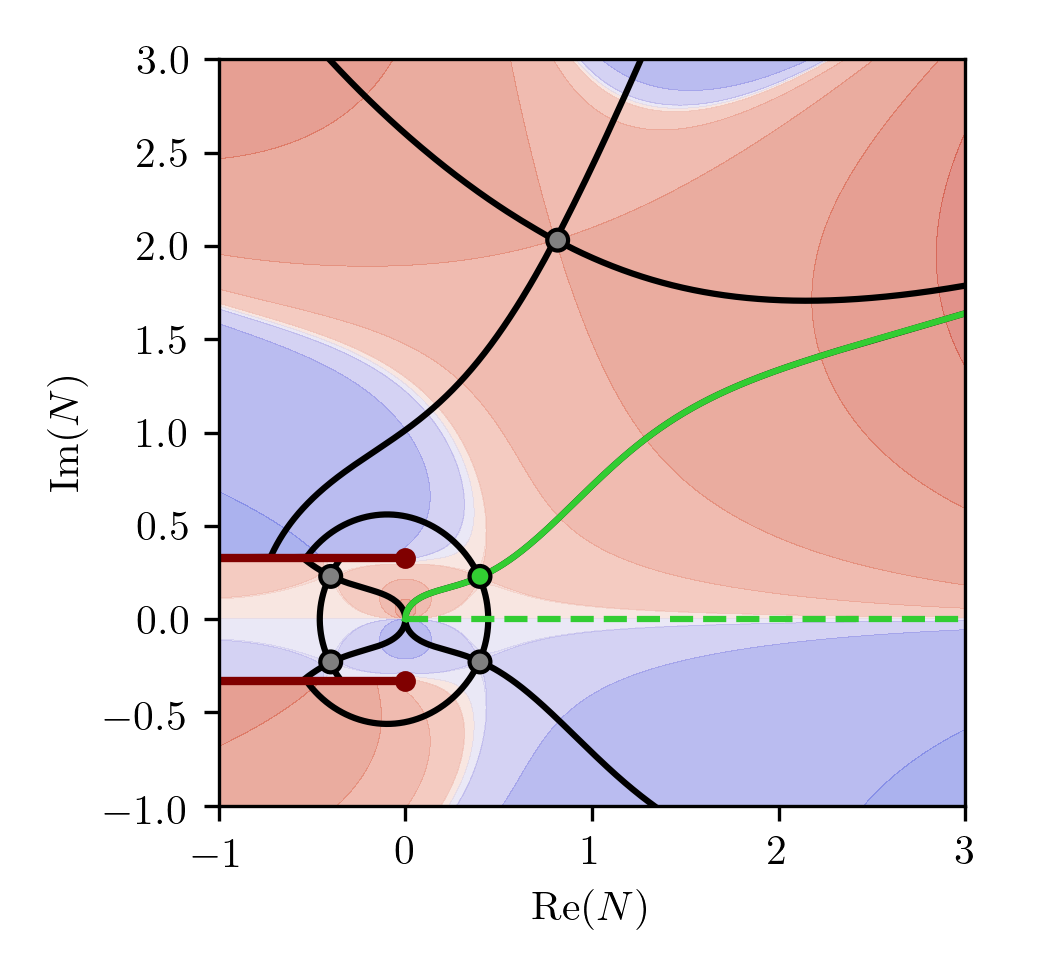}
        \subcaption{$(q_0,q_1,\tilde{n})=(0.6,1.1,1)$ and $c=-0.1$}
    \end{subfigure}
    \caption{Structure of $iS[N,q_+,n;c]$ in the complex-$N$ plane for (a) $c>0$ and (b) $c<0$. The new saddles of Eqn.~\eqref{eq:newSaddles} are far from the other saddles near the origin and can change the structure of the deformed contour for large $|N|$ (cf.\ Fig.~\ref{fig:PL_H_2}).}
    \label{fig:topPL}
\end{figure}

In the dilute gas approximation wormholes can be argued to contribute non-local terms in the action with coefficients exponentially suppressed by the wormhole action~\cite{Coleman:1988tj, Preskill:1988na, Hebecker:2018ofv}. One can choose to subsequently bring the action to a local form at the cost of introducing $\alpha$-parameters. Determining the details of this non-local action is a monumental task; indeed recently it has been proposed that such contributions should vanish~\cite{McNamara:2020uza, VanRiet:2020pcn, Schlenker:2022dyo}. In this section we show that \emph{if} such terms are present, \emph{then} they can lead to a qualitative change in the Picard-Lefschetz analysis of the transition amplitudes found above. We demonstrate this by working in the 3-form picture where the difference is most clearly seen.

Schematically, bilocal terms induced by a dilute gas of wormholes will contribute terms to the action of the form
\begin{equation}
    S \supset \frac{1}{2}\int\d[4]{x}\,\sqrt{-g}\int\d[4]{y}\,\sqrt{-g}\;C_{ij}\mathcal{O}_i(x)\mathcal{O}_j(y) \,.
\end{equation}
Restricting attention to operators of the lowest dimension, such terms will renormalize the couplings of Eqn.~\eqref{eq:gHth_action} but the largest qualitative change comes with the introduction of an effective cosmological constant (for $\mathcal{O}_i=\mathcal{O}_j=1$),
\begin{equation}
    S \supset \int\d[4]{x}\,\sqrt{-g}\,\left(-\Lambda_\text{eff}\right) \,, \qquad \Lambda_\text{eff} = \frac{c}{2\pi^2}\int\d[4]{y}\,\sqrt{-g} \,,
\end{equation}
where $|c|\sim e^{-S_\text{wh}}\ll 1$ (the factor of $2\pi^2$ will be convenient). Taking the same ansatz as in Eqn.~\eqref{eq:metricHansatz}, the transition amplitude of Eqn.~\eqref{eq:Hpath} now reads
\begin{equation}
\begin{aligned}
    Z[q_0,q_1;n] &\simeq \int_0^\infty\d{N}\int_{q(0)=q_0}^{q(1)=q_1}\mathcal{D}q\, e^{iS[N,q,n;c]} \,,\\
    S[N,q,n;c] &= 2\pi^2\int_0^1\d{t}\,\left(3N-\frac{3\dot{q}^2}{4N} - \frac{3\tilde{n}^2N}{q^2} - cN^2q\ev{q}\right) \,,
\end{aligned}
\end{equation}
where
\begin{equation}
    \ev{q} = \int_0^1\d{t'}\,q(t') \,.
\end{equation}
The equation of motion for $q$ is an integro-differential equation,
\begin{equation}
    \ddot{q} + \frac{4\tilde{n}^2N^2}{q^3} + \frac{4}{3}cN^3\ev{q} = 0 \,,
\end{equation}
which can be solved perturbatively as $q(t) = q_+(t) + c\,\delta q(t) + \mathcal{O}(c^2)$. The correction $\delta q(t)$ can be found exactly in terms of hypergeometric functions, but its detailed form will not be important for our purposes. Indeed, using the zeroth-order equations of motion the action only depends on $q_+$ at leading-order in $c$:
\begin{equation}
    S[N,q_+,n;c] = 2\pi^2\left[3N - \frac{3(q_0^2+q_1^2)}{4N}+\frac{3q_0q_1}{2N}\,f\Big(\frac{2\tilde{n}N}{q_0q_1}\Big) - cN^3\ev{q_+}^2 + \mathcal{O}(c^2)\right] \,.
\end{equation}
We have seen previously that there are several saddle points around $N\sim q_0,q_1,\tilde{n}$ (see Eqn.~\eqref{eq:saddles}). For these saddles the $c$-term is subdominant and shifts the locations of the saddle points. A qualitative change to the Picard-Lefschetz analysis comes from the appearance of new saddles points which occur for $N\gg q_0,q_1,\tilde{n}$. In this limit one has $\ev{q_+}^2\approx \frac{\pi^2\tilde{n}N}{16}$ and
\begin{equation}\label{eq:newSaddles}
    S \approx 2\pi^2\left(3N - \frac{\pi^2c\tilde{n}}{16}N^4\right) \qquad\implies\qquad N_\ast^3 \approx \frac{12}{\pi^2c\tilde{n}} \,.
\end{equation}
One can check that even with this large value of $N$ the $c$-expansion of $q(t)$ is under control:
\begin{equation*}
    \left|\frac{c\,\delta q(t)}{q_+(t)}\right| \xrightarrow{N\to\infty} \left|\frac{\pi c}{64}\, \frac{\frac{3\pi}{8}\big(1+(1-2t)^2\big) - {}_2F_1\big({-\frac{3}{2}},-\frac{1}{2},\frac{1}{2};(1-2t)^2\big)}{t(1-t)} \right| \leq \frac{\pi(3\pi-8)|c|}{128} \ll 1 \,.
\end{equation*}
If $c>0$ then $N_\ast^3>0$ and a new saddle appears on the positive real axis and always contributes as a Lorentzian saddle under the contour deformation for any $q_0,q_1$. If $c<0$ then $N_\ast^3<0$ and these new saddles \emph{never} contribute under the contour deformation. See Fig.~\ref{fig:topPL} for two representative cases.


\section{Boundary conditions \& stability}
\label{sec:stability}

Gravitational path integrals famously suffer from issues of convergence. Candidate saddle points of the Euclidean path integral should be minima so that the action at the critical point truly represents the dominant contribution from configurations near this point in field space. Saddle points (with their unstable directions) can be interpreted as mediating decay. Of course, statements of stability should only refer to gauge-invariant degrees of freedom.

In the previous sections we have restricted attention to spatially-uniform fields which obscures whether the contributing saddle points are truly stable in the appropriate sense. In order to address the question of stability we will analyze scalar perturbations around the GS wormhole in the 3-form picture, the spectrum of which depends intimately on the chosen boundary conditions; it is natural to choose Dirichlet boundary conditions for the 3-form because of flux quantization. As we saw in some detail in Sec.~\ref{sec:duality}, the duality which relates the 3-form and axion includes a correspondence between boundary conditions in the two frames: Dirichlet boundary conditions for the 3-form correspond to Neumann boundary conditions for the axion (equivalently, the Fourier transform of Dirichlet boundary conditions for the axion, in the sense discussed in Sec.~\ref{sec:duality}). Normalizable perturbations of the 3-form, namely those with finite energy for which
\begin{equation}
    \int\delta H\wedge{\star\delta H} < \infty \,,
\end{equation}
correspond, via $H\leftrightarrow{\star\d{\theta}}$, to perturbations of the axion which approach constant values at the boundaries and which have finite energy
\begin{equation}
    \int\d{\delta\theta}\wedge{\star\d{\delta\theta}} < \infty \,,
\end{equation}
even if they are not normalizable in the sense that
\begin{equation}
    \int\star(\delta\theta^2) \to \infty \,.
\end{equation}
This is natural in view of the axion's shift symmetry; a constant shift to the background field profile can be implemented with a constant perturbation which has zero energy but divergent ``norm''. Consequently the discussion of fluctuations around the GS wormhole is most transparent in the 3-form picture where Dirichlet boundary conditions are required by flux quantization and, perhaps more importantly, the criteria of normalizability and finite energy coincide.

In similar spirit to the previous sections we will remain in Lorentzian signature until absolutely necessary. We parametrize the fields as
\begin{equation}
\begin{aligned}
    \d{s^2} &= a(\eta)^2\Big\{ {-(1+2\phi)}\,\d{\eta^2} + 2\partial_iB\,\d{x^i}\d{\eta} + \big[(1-2\psi)\gamma_{ij} + 2\nabla_i\partial_jE\big]\,\d{x^i}\d{x^j} \Big\} \,,\\
    H &= \sqrt{6}\,\tilde{n}\left[(1+s)\,\vol_3 + \d{\eta}\wedge\left(\frac{1}{2}\sqrt{\gamma}\gamma^{ij}\epsilon_{jkl}\partial_iw\,\d{x^k}\wedge\d{x^l}\right)\right] \,,
\end{aligned}
\end{equation}
where $\gamma_{ij}$ is the (fixed) round metric on $S^3$ and $\nabla$ the corresponding covariant derivative. Note that we are using conformal time $\eta$, so now $\dot{A}=\de{A}{\eta}$. It will be useful to introduce $\mathcal{H}=\frac{\dot{a}}{a}$, in terms of which the zeroth-order Einstein equations amount to
\begin{equation}\label{eq:hubbleEOM}
    1+\mathcal{H}^2 = \frac{\tilde{n}^2}{a^4} > 0 \,.
\end{equation}
Returning to the action of Eqn.~\eqref{eq:gHth_action} and using the above parametrization results in the following quadratic action for the perturbations,
\begin{equation}
\begin{aligned}
    S_2 &= \int\d{\eta}\d[3]{x}\,\sqrt{\gamma}\,a^2\Big\{ {-3}\big(\dot{\psi} + \mathcal{H}\phi\big)^2 + \big(B-\dot{E}\big)\Delta\big(B-\dot{E}\big) - 2\big(\dot{\psi}+\mathcal{H}\phi\big)\Delta\big(B-\dot{E}\big)\\
    &\qquad\qquad - 3\big(1+\mathcal{H}^2\big)\big[(\phi+3\psi-\Delta E+s)^2 -\phi^2+ (B-w)\Delta(B-w)\big]\\
    &\qquad\qquad + (2\phi-\psi)(\Delta+3)\psi\Big\} + \sqrt{6}\tilde{n}\int\d{\eta}\d[3]{x}\,\sqrt{\gamma} \,(\dot{s}-\Delta w)\theta \,,
\end{aligned}
\end{equation}
where we have used integration by parts on $S^3$ liberally. The individual perturbations are not gauge-invariant; under a diffeomorphism $\xi=\zeta^0\partial_\eta + \gamma^{ij}(\partial_i\zeta)\partial_j$ parametrized by the two scalar functions $\zeta^0,\zeta$ the perturbations transform according to $\mathcal{L}_\xi g$ and $\mathcal{L}_\xi H$:
\begin{equation}
\begin{aligned}
    \delta_\xi\phi &= \dot{\zeta}^0+\mathcal{H}\zeta^0 \,, & \qquad \delta_\xi B &= -\zeta^0+\dot{\zeta} \,, & \qquad \delta_\xi s &= \Delta\zeta \,,\\
    \delta_\xi\psi &= -\mathcal{H}\zeta^0 \,, & \delta_\xi E &= \zeta \,, & \delta_\xi w &= \dot{\zeta} \,.
\end{aligned}
\end{equation}
Physically meaningful statements can only be made about linear combinations of perturbations which are gauge-invariant.

\begin{figure}[t]
    \centering
    \includegraphics[width=\textwidth]{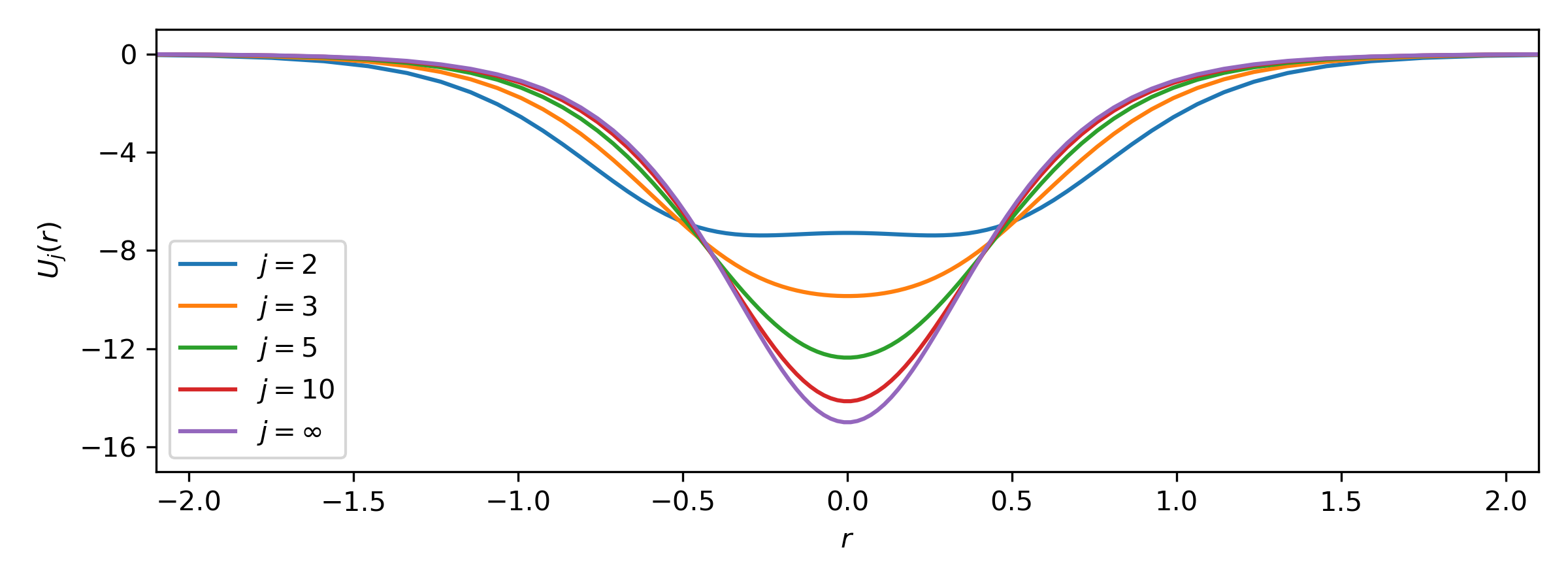}
    \caption{The potentials $U_j$ for the Schr\"odinger problem of Eqn.~\eqref{eq:Schrodinger}.}
    \label{fig:potentials}
\end{figure}

To proceed it is useful to reduce to 1D by writing all fields in terms of hyperspherical harmonics, e.g.
\begin{equation}
    \phi(\eta,x^i) = \sum_{j\geq0} \phi_j(\eta)Y_j(x^i)
\end{equation}
where
\begin{equation}
    \Delta Y_j = -\lambda_jY_j \,, \qquad \int_{S^3}\d[3]{x}\,\sqrt{\gamma}\,Y_jY_{j'} = \delta_{jj'} \,.
\end{equation}
The degeneracy of the eigenvalue $\lambda_j=j(j+2)\in\{0,3,8,15,\ldots\}$ is $(j+1)^2$; we suppress labels which distinguish degenerate states for simplicity. The action decomposes into sectors labeled by the integer $j$:
\begin{equation}
\begin{aligned}
    S_2 &= \int\d{\eta}\,\sum_{j\geq0}\mathcal{L}_j \,,\\
    \mathcal{L}_j &= a^2\Big\{ {-3}\big(\dot{\psi}_j + \mathcal{H}\phi_j\big)^2 - \lambda_j\big(B_j-\dot{E}_j\big)^2 + 2\lambda_j\big(\dot{\psi}_j+\mathcal{H}\phi_j\big)\big(B_j-\dot{E}_j\big)\\
    &\qquad\qquad - 3\big(1+\mathcal{H}^2\big)\big[(\phi_j+3\psi_j + \lambda_jE_j+s_j)^2 -\phi_j^2 - \lambda_j(B_j-w_j)^2\big]\\
    &\qquad\qquad - (\lambda_j-3)(2\phi_j-\psi_j)\psi_j \Big\} + \sqrt{6}\tilde{n}(\dot{s}_j+\lambda_jw_j)\theta_j \,.
\end{aligned}
\end{equation}
For now let us focus on the sectors with $j\geq2$ ($\lambda_j\geq8$) which can all be treated simultaneously. Performing the path integral over the Lagrange multiplier $\theta_j$ results in a $\delta$-function imposing the (gauge-invariant) condition $\dot{s}_j+\lambda_jw_j=0$. Using this to eliminate $w_j$ leads to
\begin{equation}
\begin{aligned}
    \mathcal{L}_j &= a^2\Big\{ {-3}\big(\dot{\psi}_j + \mathcal{H}\phi_j\big)^2 - \lambda_j\big(B_j-\dot{E}_j\big)^2 + 2\lambda_j\big(\dot{\psi}_j+\mathcal{H}\phi_j\big)\big(B_j-\dot{E}_j\big)\\
    &\qquad\qquad - 3\big(1+\mathcal{H}^2\big)\big[(\phi_j+3\psi_j + \lambda_jE_j+s_j)^2 -\phi_j^2 - \lambda_j^{-1}(\dot{s}_j+\lambda_jB_j)^2\big]\\
    &\qquad\qquad - (\lambda_j-3)(2\phi_j-\psi_j)\psi_j \Big\} \,.
\end{aligned}
\end{equation}
All of $\psi_j,E_j,s_j$ are dynamical and have conjugate momenta given by
\begin{equation}
\begin{aligned}
    \Pi_j^\psi &= 2a^2\Big[ {-3}\big(\dot{\psi}_j+\mathcal{H}\phi_j\big) + \lambda_j\big(B_j-\dot{E}_j\big) \Big] \,,\\
    \Pi_j^E &= 2a^2\Big[ {-\lambda_j}\big(\dot{\psi}_j+\mathcal{H}\phi_j\big) + \lambda_j\big(B_j-\dot{E}_j\big) \Big] \,,\\
    \Pi_j^s &= 6\lambda_j^{-1}a^2\big(1+\mathcal{H}^2\big)(\dot{s}_j+\lambda_jB_j) \,.
\end{aligned}
\end{equation}
In terms of these we may write the Lagrangian in first-order form:
\begin{equation}
\begin{aligned}
    \mathcal{L}_j &= \Pi_j^\psi\dot{\psi}_j + \Pi_j^E\dot{E}_j + \Pi_j^s\dot{s}_j - (\Pi_j^E-\lambda_j\Pi_j^s)B_j\\
    &\qquad + \big[\mathcal{H}\Pi_j^\psi - 6a^2\big(1+\mathcal{H}^2\big)(3\psi_j+\lambda_jE_j+s_j) - 2a^2(\lambda_j-3)\psi_j\big]\phi_j\\
    &\qquad + a^{-2}\left[ -\frac{(\Pi_j^\psi)^2}{4(\lambda_j-3)} + \frac{\Pi_j^\psi\Pi_j^E}{2(\lambda_j-3)} - \frac{3(\Pi_j^E)^2}{4\lambda_j(\lambda_j-3)} - \frac{\lambda_j(\Pi_j^s)^2}{12(1+\mathcal{H}^2)} \right]\\
    &\qquad - a^2\Big[ 3\big(1+\mathcal{H}^2\big) (3\psi_j+\lambda_jE_j+s_j)^2 -(\lambda_j-3)\psi_j^2 \Big] \,.
\end{aligned}
\end{equation}
This is linear in both of the non-dynamical fields, $\phi_j$ and $B_j$, and performing the path integral over them produces two more (gauge-invariant) $\delta$-function constraints. Using these to integrate out $\Pi_j^\psi$ and $\Pi_j^E$ gives
\begin{equation}
\begin{aligned}
    \mathcal{L}_j &= \Pi_j^S\dot{S}_j - \frac{\lambda_j}{12a^2}\left(\frac{9}{\lambda_j-3} + \frac{1}{1+\mathcal{H}^2} \right)(\Pi_j^S)^2\\
    &\qquad + \frac{3\lambda_j(1+\mathcal{H}^2)}{(\lambda_j-3)\mathcal{H}}\Pi_j^SS_j - \frac{3a^2(1+\mathcal{H}^2)^2}{\mathcal{H}^2}\left(\frac{\lambda_j}{\lambda_j-3} - \frac{1}{1+\mathcal{H}^2}\right)S^2
\end{aligned}
\end{equation}
(plus a total derivative involving $\psi_j$), where we have introduced the gauge-invariant field $S_j$ and its gauge-invariant conjugate momentum $\Pi_j^S$ defined by
\begin{equation}
\begin{aligned}
    S_j &= s_j + \lambda_jE_j \,,\\
    \Pi_j^S &= \Pi_j^s - 6a^2(1+\mathcal{H}^2)\mathcal{H}^{-1}\psi \,.
\end{aligned}
\end{equation}
The fields $\psi_j,E_j$ no longer appear; we can interpret the path integral over these as giving the gauge-orbit volume. Finally, integrating out $\Pi_j^S$ returns the Lagrangian to second-order form, now for the sole physical, gauge-invariant scalar perturbation:
\begin{equation}
    \mathcal{L}_j = \frac{3a^2}{\lambda_j\big(\frac{9}{\lambda_j-3}+\frac{1}{1+\mathcal{H}^2}\big)}\left[ \dot{S}_j^2 + \frac{6\lambda_j\big(1+\mathcal{H}^2\big)}{(\lambda_j-3)\mathcal{H}}S_j\dot{S}_j - \lambda_j\left(\frac{\lambda_j-9}{\lambda_j-3}\big(1+\mathcal{H}^2\big) - 1\right)S_j^2 \right] \,.
\end{equation}
A similar analysis for $j=0$ and $j=1$ reveals that these two sectors are pure-gauge and as such do not correspond to any physical perturbations.\footnote{For $j=0$ the Lagrange multiplier imposes $\dot{s}_0=0$, so $s_0=0$ for the given boundary conditions. Additionally, all terms involving $B_0,E_0,w_0$ vanish and of the two remaining fields, $\phi_0$ and $\psi_0$, only $\psi_0$ is dynamical. Transforming to gauge-invariant variables and integrating out the remaining non-dynamical field shows that $\mathcal{L}_0$ is a total derivative. Similarly, for $j=1$ one can replace $w_1\to-\dot{s}_1$ using the constraint to find that there are two dynamical fields, $\psi_1+E_1$ and $s_1$, and two non-dynamical fields, $\phi_1$ and $B_1$ (the linear combination $\psi_1-E_1$ does not appear). Integrating out the non-dynamical fields imposes two gauge-invariant constraints which result in $\mathcal{L}_1$ being a total derivative.} In particular, there is no conformal factor problem in the homogeneous ($j=0$) sector (the same observation was made in~\cite{Hertog:2018kbz}). This can be understood as a consequence of the zeroth-order Einstein equations of Eqn.~\eqref{eq:hubbleEOM} being first-order. In terms of the canonically normalized field
\begin{equation}
    Q_j = \sqrt{\frac{6a^2}{\lambda_j\big(\frac{9}{\lambda_j-3}+\frac{1}{1+\mathcal{H}^2}\big)}}\;S_j
\end{equation}
(note that the ratio in the square-root is positive since $1+\mathcal{H}^2>0$ by Eqn.~\eqref{eq:hubbleEOM} and $\lambda_j>3$) the Lagrangian is
\begin{align}
    \mathcal{L}_j &= \frac{1}{2}\dot{Q}_j^2 - \frac{1}{2}\big[U_j + (\lambda_j+1)\big] Q_j^2 + \de{}{\eta}\left[ \left(\frac{\mathcal{H}}{2}\frac{9(1+\mathcal{H}^2)-(\lambda_j-3)}{9(1+\mathcal{H}^2)+(\lambda_j-3)}+\frac{3\lambda_j(1+\mathcal{H}^2)}{2\mathcal{H}(\lambda_j-3)}\right)Q_j^2 \right] \,, \notag\\
    U_j &= -\big(1+\mathcal{H}^2\big)\left(4(\lambda_j+6)\,\frac{9(1+\mathcal{H}^2)+4(\lambda_j-3)}{[9(1+\mathcal{H}^2)+(\lambda_j-3)]^2} - 1 \right) \,.
\end{align}
At this point there is no obstacle to performing the standard Wick rotation to Euclidean signature, $\eta\to-ir$. Under this rotation it is useful to introduce $A'\equiv\de{A}{r}$ and $\mathcal{H}_\text{E}(r) = -i\mathcal{H}(ir) = \frac{a'(r)}{a(r)}$ for clarity; the replacement $\mathcal{H}^2\to-\mathcal{H}_\text{E}^2$ should be made in both $Q_j$ and $U_j$. The equation of motion Eqn.~\eqref{eq:hubbleEOM} now reads
\begin{equation}
    1 \geq 1 - \mathcal{H}_\text{E}^2 = \frac{\tilde{n}^2}{a^4} > 0
\end{equation}
and has solution
\begin{equation}
    a(r) = \sqrt{\tilde{n}\cosh{(2r)}} \,, \qquad \mathcal{H}_\text{E}(r) = \tanh{(2r)} \,.
\end{equation}
This is the GS wormhole; the wormhole throat is at $r=0$ and the two asymptotically flat regions are $r\to\pm\infty$. The Euclidean action becomes
\begin{align}
    S_\text{E} &= \int\d{r}\sum_{j\geq2}\left\{\frac{1}{2}Q_j\Big[\hat{\mathcal{O}}_j + (\lambda_j+1)\Big]Q_j + G_j\right\} \,, \notag\\
    \hat{\mathcal{O}}_j &= -\de[2]{}{r} + U_j \,, \\
    G_j &= \frac{1}{2}(Q_jQ_j')' + \mathcal{H}_\text{E}\left( \frac{9(1-\mathcal{H}_\text{E}^2)-(\lambda_j-3)}{9(1-\mathcal{H}_\text{E}^2)+(\lambda_j-3)}-\frac{3\lambda_j(1-\mathcal{H}_\text{E}^2)}{(\lambda_j-3)\mathcal{H}_\text{E}^2}  \right)Q_jQ_j' \notag\\
    &\quad + (1-\mathcal{H}_\text{E}^2)\left( \frac{[9(1-\mathcal{H}_\text{E}^2)+2(\lambda_j-3)]^2-(\lambda_j-3)(5\lambda_j+21)}{[9(1-\mathcal{H}_\text{E}^2)+(\lambda_j-3)]^2} + \frac{3\lambda_j(1+\mathcal{H}_\text{E}^2)}{(\lambda_j-3)\mathcal{H}_\text{E}^2} \right)Q_j^2 \notag\\
    &= \frac{1}{2}\left[Q_jQ_j' + \mathcal{H}_\text{E}\left(\frac{9(1-\mathcal{H}_\text{E}^2)-(\lambda_j-3)}{9(1-\mathcal{H}_\text{E}^2)+(\lambda_j-3)}-\frac{3\lambda_j(1-\mathcal{H}_\text{E}^2)}{(\lambda_j-3)\mathcal{H}_\text{E}^2}\right)Q_j^2\right]' \,, \notag
\end{align}
where the constant $\lambda_j+1$ has been stripped off in defining $U_j$ so that $U_j\to0$ in the two asymptotic regions (see Fig.~\ref{fig:potentials}); note, however, that $U_j+(\lambda_j+1)>0$ everywhere. The term $G_j$ is a total derivative and has no effect on the equations of motion for $Q_j$. Nevertheless, once solutions have been found one has to check that $S_\text{E}$ is positive definite when including the contributions from integrating over $G_j$.

Finding the spectrum of scalar fluctuations amounts to finding eigenvalues of the fluctuation operators $\hat{\mathcal{O}}_j$ given appropriate boundary conditions. For each $j$ this is a standard Schr\"odinger-type problem with potential $U_j$:
\begin{equation}\label{eq:Schrodinger}
    \hat{\mathcal{O}}_jQ_j^{(k)} = \left(-\de[2]{}{r} + U_j\right)Q_j^{(k)} = \omega_j^{(k)}Q_j^{(k)} \,.
\end{equation}
What boundary conditions should be required for the canonically normalized fields? Although physical perturbations should vanish at the boundaries, the normalization of $Q_j$,
\begin{equation}
    Q_j = \sqrt{\frac{6\tilde{n}\cosh{(2r)}}{\lambda_j\big[\frac{9}{\lambda_j-3}+\cosh^2{(2r)}\big]}}\;S_j \quad\xrightarrow{|r|\to\infty}\quad \sqrt{\frac{12\tilde{n}}{\lambda_j}}\;e^{-|r|}S_j \,,
\end{equation}
shows that we should insist that $Q_j\to0$ \emph{faster} than $e^{-|r|}$ for $|r|\to\infty$ in order that the corresponding perturbation $S_j$ goes to zero in both asymptotic regions. Any bound states of energy \emph{strictly less than} ${-1}$ will have exponential tails in the classically forbidden region which die off quickly enough. Note also that the normalizing factor in $Q_j$ is well-behaved near $r=0$, so no additional conditions need be imposed there.

\begin{figure}[t]
    \centering
    \includegraphics[width=\textwidth]{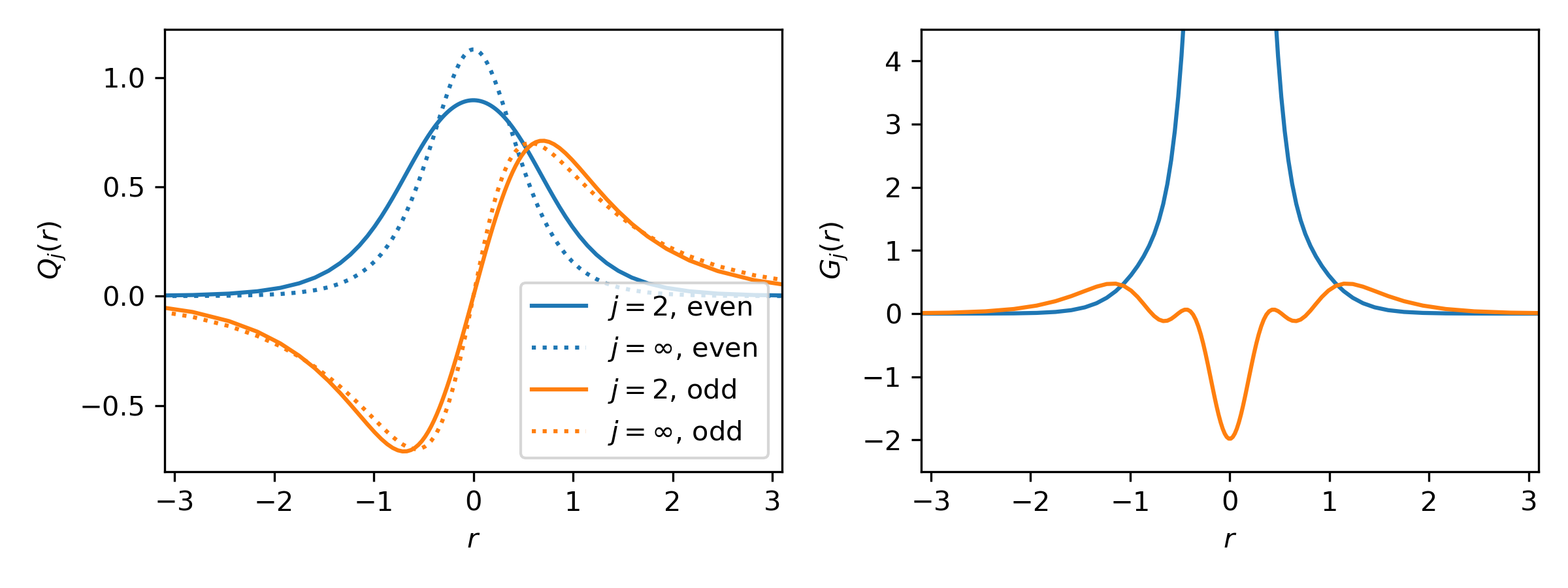}
    \caption{(Left) Even and odd bound states for $j=2$ and $j\to\infty$. (Right) The functions $G_j(r)$ for the even and odd $j=2$ bound states. For the even bound state the integral over $G_j$ diverges while for the odd bound state it vanishes.}
    \label{fig:eigen}
\end{figure}

For $j\gg1$ the functions $U_j$ approach a universal, P\"oschl--Teller form,
\begin{equation}
    U_\infty = -\frac{15}{\cosh^2{(2r)}} \,,
\end{equation}
for which there are only two bound states, one even in $r$ and the other odd:
\begin{equation}
\begin{aligned}
    Q_\infty^{\text{(e)}} &= \frac{2}{\sqrt{\pi}}\frac{1}{[\cosh{(2r)}]^{3/2}} \,, & \qquad \omega_\infty^{\text{(e)}} &= -9 \,,\\
    Q_\infty^{\text{(o)}} &= \frac{2}{\sqrt{\pi}}\frac{\sinh{(2r)}}{[\cosh{(2r)}]^{3/2}} \,, & \omega_\infty^{\text{(o)}} &= -1 \,.
\end{aligned}
\end{equation}
One can check numerically that exactly two bound states, one even and one odd, exist for all $j\geq2$. These bound states, $Q_j^\text{(e)}$ and $Q_j^\text{(o)}$, are perturbed versions of their $j\to\infty$ counterparts: see Fig.~\ref{fig:eigen}. Although $Q_\infty^{\text{(o)}}$ marginally violates the required fall-off conditions, one can use standard time-independent perturbation theory to show that
\begin{equation}\label{eq:lambdaapprox}
\begin{aligned}
    \omega_j^\text{(e)} &= -9 + \frac{99}{2\lambda_j} + \mathcal{O}(\lambda_j^{-2}) \,,\\
    \omega_j^\text{(o)} &= -1 - \frac{9}{2\lambda_j} + \mathcal{O}(\lambda_j^{-2}) \,,
\end{aligned}
\end{equation}
so that the required fall-off conditions at $|r|\to\infty$ are in fact satisfied by the odd bound states for finite $j$. In evaluating the action for these states one finds that $G_j$ diverges near $r=0$ as $G_j\sim \frac{1}{r^2}>0$ for the even bound states $Q_j^{\text{(e)}}$; the Euclidean action diverges $S_\text{E}\to+\infty$ and these states are discarded. In contrast, the odd states survive as finite-action perturbations since $G_j$ is everywhere finite and integrates to zero: see Fig.~\ref{fig:eigen}. All told there is (up to the $(j+1)^2$ degeneracy) a unique admissible scalar perturbation for each $j\geq2$ with
\begin{equation}
    S_\text{E} = \frac{1}{2}\int\d{r}\,Q_j\Big[\hat{\mathcal{O}}_j+(\lambda_j+1)\Big]Q_j = \frac{1}{2}\big(\omega_j^\text{(o)} + \lambda_j+1\big) > 0 \,.
\end{equation}
The low-lying spectrum is found using finite-difference techniques and presented in Tab.~\ref{tab:spectrum}. We conclude that the GS wormhole is perturbatively stable.

\begin{table}[t]
    \centering
    \begin{tabular}{c|ccc||c|ccc}
        $j$ & $-1-\frac{9}{2\lambda_j}$ & $\omega_j^\text{(o)}$ & $\omega_j^\text{(o)}+\lambda_j+1$ & $j$ & $-1-\frac{9}{2\lambda_j}$ & $\omega_j^\text{(o)}$ & $\omega_j^\text{(o)}+\lambda_j+1$\\ \hline
        2 & $-1.5625$ & $-1.5335$ & $\phantom{0}7.4665$ & 6 & $-1.0938$ & $-1.0921$ & $47.9079$\\
        3 & $-1.3000$ & $-1.2873$ & $14.7127$ & 7 & $-1.0714$ & $-1.0705$ & $62.9295$\\
        4 & $-1.1875$ & $-1.1817$ & $23.8183$ & 8 & $-1.0563$ & $-1.0556$ & $79.9444$\\
        5 & $-1.1286$ & $-1.1256$ & $34.8744$ & 9 & $-1.0455$ & $-1.0450$ & $98.9550$
    \end{tabular}
    \caption{Low-lying spectrum of finite-action scalar perturbations, along with the approximation of Eqn.~\eqref{eq:lambdaapprox}.}
    \label{tab:spectrum}
\end{table}


\section{Discussion}
\label{sec:disc}

In this paper, we have explored several aspects of the dual axion-gravity and gravity$+$3-form theories directly with the Lorentzian path integral. Using Picard-Lefschetz theory we are able to identify which subset of (complex) saddle points contribute to the Lorentzian path integrals computing simple transition amplitudes. As the boundary values are adjusted the interpretation in terms of Lorentzian-time evolution vs.\ Euclidean wormhole changes, but nevertheless they may all be treated democratically in the complex-$N$ plane. All four saddle points are compatible with the K-S criterion~\cite{Kontsevich2021} but only ever one or two are selected by the contour deformation. It would be interesting to understand if another principle could be used to omit the saddles that do not contribute according to Picard-Lefschetz theory.

We considered scalar perturbations to the GS wormhole in the 3-form picture, where it was found that (i) there is no conformal factor problem in the homogeneous sector and (ii) there are no negative modes amongst the physical perturbations of nonzero angular momentum. The perturbative stability of the GS wormhole presents several puzzles. For one, our results disagree with the conclusion of~\cite{Hertog:2018kbz} which analyzed stability in the axion picture and found that there are infinitely many negative modes. Here the application of boundary conditions is more straightforward since both metric and 3-form scalar perturbations are both subject to Dirichlet boundary conditions, whereas the axion is subject to Neumann boundary conditions and the gauge-invariant field chosen in~\cite{Hertog:2018kbz} mixes metric and axion perturbations. We have also seen the importance of keeping all boundary terms for the gauge-invariant modes $Q_j$, namely in ruling out the even-parity perturbations for which the total derivative terms $G_j$ are not integrable.

Our findings contribute to the long-standing debate on whether Euclidean wormholes can be embedded into string theory. Previous studies found negative modes \cite{Hertog:2018kbz,Rubakov:1996cn,Kim:2003js} suggesting that axionic Euclidean wormholes are not saddle points of the gravitational path integral. Here we showed that with appropriate boundary conditions, the spectrum of quadratic fluctuations contains only modes with positive eigenvalues. This suggests the interpretation that rather than mediating decay these wormholes compute the non-perturbative energy splitting for the degenerate vacua with flux $\pm n$ (note that for $n=0$ this 2-fold degeneracy disappears, as does the GS wormhole).
While we have analyzed the stability of axionic wormholes in asymptotically-flat space, the conclusion of stability may carry over to AdS since the perturbations we find are localized to the wormhole's throat whose size is much less than the AdS curvature.
These Euclidean wormholes, if embeddable into AdS compactifications, pose a puzzle for AdS/CFT as they seem to jeopardize factorization of the two boundary CFTs.
Given this potential tension with AdS/CFT the embeddability of these solutions into string theory is called into question. There are a few logically possible resolutions:
\begin{enumerate}[label=\roman*)]
    \item The stability of axionic Euclidean wormholes changes qualitatively with the inclusion of a dilaton (which always comes along for the ride with an axion) or nonzero cosmological constant. It would be interesting to extend our current work to analyze the spectrum of gauge invariant perturbations of such axio-dilatonic Euclidean wormholes. It may also be that these wormholes are non-perturbatively unstable to brane nucleation, such as was found in~\cite{Marolf:2021kjc}.
    \item The existence of a regular wormhole solution requires finding a long-enough timelike geodesic in the scalar moduli space \cite{Arkani-Hamed:2007cpn}. This in turn implies conditions on the dilaton couplings \cite{Arkani-Hamed:2007cpn, Astesiano:2022qba}. It may happen that such couplings are not realizable in controlled limits of string theory. These embeddability criteria are reminiscent of swampland criteria such as the distance conjecture \cite{Ooguri:2006in} which limits the field range for a single effective field theory description to be valid (albeit for positive definite field space metric). Our Lorentzian, 3-form approach may allow us to formulate these embeddability criteria with a more familiar positive-definite moduli space metric.
    \item The axio-dilatonic Euclidean wormholes, if shown to be stable and embeddable into AdS compactifications, suggest a more general interpretation of the AdS/CFT duality in $D\geq3$, for example involving an ensemble-average. This would be at odds with~\cite{McNamara:2020uza} and~\cite{Schlenker:2022dyo} (see also~\cite{Heckman:2021vzx}).
\end{enumerate}
Further studies along these lines may teach us useful lessons about quantum gravity. We plan to investigate these possibilities in the future.


\acknowledgments

We thank Thomas Van Riet for useful discussions. The work of GL and GS is supported in part by the DOE grant DE-SC0017647.


\newpage
\bibliographystyle{utphys}
\bibliography{wormholePL}

\end{document}